\documentclass[aps,twocolumn,pre,nofootinbib]{revtex4}
\usepackage[utf8x]{inputenc}
\usepackage{graphicx}
\usepackage{latexsym}
\usepackage{amsmath}
\usepackage{amssymb}
\usepackage{bm}

\newcommand{\bee}{\begin{eqnarray}}
\newcommand{\eee}{\end{eqnarray}}
\newcommand{\ba}{\begin{array}}
\newcommand{\ea}{\end{array}}
\newcommand{\bc}{\begin{center}}
\newcommand{\ec}{\end{center}}
\newcommand{\bi}{\begin{itemize}}
\newcommand{\ei}{\end{itemize}}
\newcommand{\br}{\mbox{\boldmath $r$}}
\newcommand{\oOmega}{\mbox{\boldmath $\Omega$}}

\newcommand{\m}{\mbox{\boldmath $\mu$}}

\newcommand{\nnabla}{\mbox{\boldmath $\nabla$}}
 	
\begin{document}
\title{Hydrodynamic radius approximation for spherical particles suspended in a viscous fluid: influence of particle internal structure and boundary}
\author{Bogdan Cichocki}
 \affiliation{Institute of Theoretical Physics, Faculty of Physics, University of Warsaw, Ho\.za 69,
  00-681 Warsaw, Poland}

\author{Maria L. Ekiel-Je\.zewska}
\email{mekiel@ippt.pan.pl}
 \affiliation{Institute of Fundamental Technological Research,
             Polish Academy of Sciences, Pawi\'nskiego 5B, 02-106 Warsaw, Poland}

\author{Eligiusz Wajnryb}
 \affiliation{Institute of Fundamental Technological Research,
             Polish Academy of Sciences, Pawi\'nskiego 5B, 02-106 Warsaw, Poland}

\date{\today}
\begin{abstract}
Systems of spherical particles moving in Stokes flow are studied for a different particle internal structure and boundaries, including the Navier-slip model. It is shown that their hydrodynamic interactions are well described by treating them as solid spheres of smaller  hydrodynamic radii, which can be determined from measured single-particle diffusion or intrinsic viscosity coefficients. Effective dynamics of suspensions made of such particles is quite accurately described by mobility coefficients of the solid particles with the  hydrodynamic radii, averaged with the unchanged direct interactions between the particles.
\end{abstract}

\maketitle
\section{Introduction}
Hydrodynamic effects of surface layers or surface roughness on micro or nanoparticles are important for dynamics and rheology of systems, which contain many such particles. Examples are micro and nanogels, which has been recently investigated in many contexts, especially with the perspective of using them as ``containers'' which can be used in the processes of drug or protein delivery~\cite{Pelton,x}. 

Anderson et al. analyzed theoretically hydrodynamic thickness of thin polymer layers at solid-liquid surfaces~\cite{Anderson1}, and determined the effect of polymer layers adsorbed to colloidal particles on their basic hydrodynamic properties~\cite{Anderson2}.

In this paper, the results from Refs.~\onlinecite{Anderson1} and~\onlinecite{Anderson2} are generalized for a wide class of the boundary conditions at the particle surfaces, and for {\it arbitrary} ambient flows, including flows generated by other suspension particles. The goal is to show that hydrodynamic properties of spherical particles with a different internal structure and boundaries can be well-approximated only in terms of their  hydrodynamic radii and direct interactions, such as e.g. the no-overlap condition or electrostatic repulsion. 

The concept of the  hydrodynamic radius is widely used in the literature as the basic feature deduced from single-particle experiments, e.g. the DLS measurements of the translational self-diffusion coefficients of dilute suspensions, combined with the Stokes-Einstein relation~\cite{Dhont}, or viscometric measurements of the intrinsic viscosity, supplemented by the Einstein's theory~\cite{SS}. This concept is often also applied to non-spherical particles~\cite{AESWW}. However, in this work we analyze the concept of the  hydrodynamic radius only in the context of spherical particles, and we show how to apply it to account for the hydrodynamics of particles with a different internal structure and boundaries.  

The plan of the paper is the following. With the ultimate goal to understand hydrodynamics of spherical particles, we start in Sec.~\ref{2} from analyzing a simplified problem: a fluid separated from a permeable medium by a planar interface or bounded by a flat rough solid surface. It is shown that the fluid flow can be approximated by the flow of the fluid bounded by an effective planar smooth solid surface. In Sec.~\ref{3}, the system of isolated spherical particles, permeable or with rough surfaces, is considered.

The  hydrodynamic radius model is constructed and justified, based on the multipole method of solving the Stokes equations. In Sec.~\ref{many}, hydrodynamic interactions between many particles are analyzed. The method of reflections (scatterings) is described, and the Rotne-Prager approximation is derived for the mobility of many particles described by the  hydrodynamic radius model.  In Sec.~\ref{direct}, the dynamics of suspensions is analyzed, and the role of direct interactions 
in averaging the effective mobility coefficients is discussed. Conclusions are presented in Sec.~\ref{5}. 

\section{Basic concept: shear flow close to a flat interface or solid wall}\label{2}
\begin{figure}[h] 
\bc
\includegraphics[width=7cm]{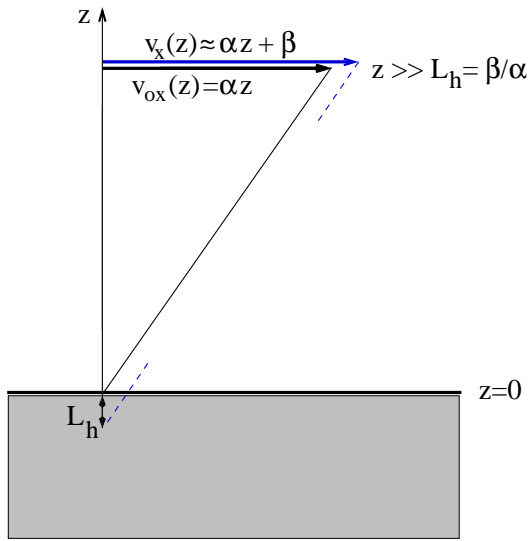}\ec
\caption{A simplified system: fluid bounded by an interface. The slip length $L_h$ determines the position of the effective hydrodynamic boundary.}\label{f11}
\end{figure}

\begin{figure}[h] 
\bc
\vspace{-0.1cm}
\includegraphics[width=5cm]{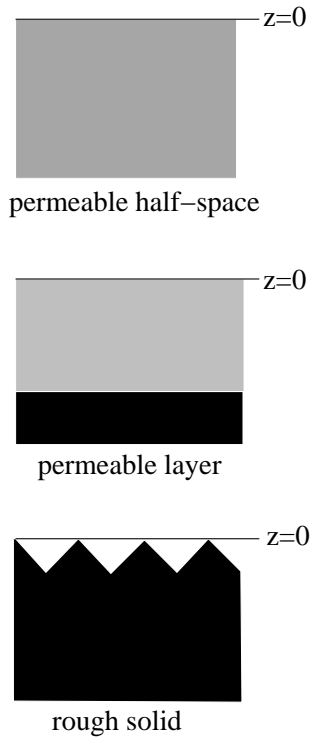}\ec \vspace{-0.3cm}
\caption{Examples of fluid boundaries, which are hydrodynamically well-approximated by an effective flat solid surface. }\label{f2}
\end{figure}

Before focusing on hydrodynamics of spherical particles made of a given material, we analyze an auxiliary problem of a flat boundary between such a material and the fluid. 
We assume that the half-space  $z > 0$ is filled with a viscous fluid, which at $z=0$ is separated from a certain medium, and moves as a shear flow when $z \rightarrow \infty$, as shown in Fig. \ref{f11}. 
The plane $z=0$ represents the boundary of a  permeable medium, which fills the half-space or forms a layer, or 
a flat surface which touches the highest peaks of the  random or regular (periodic) solid asperities, as illustrated in Fig.~\ref{f2}.

The fluid velocity ${\bf v}(\br)$ and pressure $p(\br)$ in the half-space $z \ge 0$ satisfy the Stokes equations,
\bee
\eta {\bf \nnabla}^2 
{\bf v}(\br) -{\bf \nnabla} p(\br) &=& 0, \label{St1}\\
{\bf \nnabla} \cdot {\bf v}(\br) &=& 0,\label{St}
\eee
 with the following boundary conditions,
\bee
{\bf v}(z) \sim {\bf v}_0(z)=\alpha z {\bf e}_x\hspace{0.5cm} &\mbox{for}& \hspace{0.5cm} z \rightarrow \infty,\eee and
$
{\bf v}(z) \rightarrow 0$  at the rough solid surface,
or
continuity of the fluid velocity and tangential stress at the interface $z=0$ of a permeable medium, with ${\bf v}(z) \rightarrow 0$ at $z \rightarrow - \infty$ or at the other boundary of the permeable layer. In particular, such boundary conditions exclude fluid-fluid boundaries. 

For $z \rightarrow + \infty$, the solution has the form,
\bee
{\bf v}(z) \approx \alpha z  {\bf e}_x[1 + L_h/z + {o} (L_h/z)],
\eee
where $L_h$ depends on the medium at $z < 0$. 

For example, consider a uniformly permeable medium with the hydrodynamic screening length $\kappa^{-1}$, described by the Brinkman-Debye-Bueche equations~\cite{Brinkman:47,DebyeBueche:48},
\bee
\eta ({\bf \nnabla}^2 - \kappa^2)
{\bf v}(\br) -{\bf \nnabla} p(\br) &=& 0, \label{BDB1}\\
{\bf \nnabla} \cdot {\bf v}(\br) &=& 0.\label{BDB2}
\eee 
If the medium fills the half-space $z < 0$, then the solution of the above problem gives 
\bee
L_h=\kappa^{-1}.\label{per}
\eee 
If the medium forms a layer, with $-\delta < z < 0$, then 
\bee
L_h=\kappa^{-1} \tanh (\kappa \delta),\label{cs}
\eee 
see Eq.~(20a) in Ref.~\onlinecite{Anderson1}. The explicit form of $L_h$ has been also derived for a layer made of non-uniformly permeable medium, with the exponential profile of the segment density of the adsorbed polymer, see Eq. (20b) in Ref.~\onlinecite{Anderson1}.

For the Navier boundary conditions at the solid surface $z=0$~\cite{Navier_1813} (also called the stick-slip or mixed stick-slip~\cite{Felderhof76}), given in terms of the stress tensor 
$\sigma_{ij}=\eta(\partial_i v_j + \partial_jv_i)-\delta_{ij}p$ by the following expressions, 
\bee
v_i&=&\dfrac{\lambda}{\eta} \sigma_{iz}, \;\;\;\mbox{ for } i=x,y,\\
 v_z&=&0, \eee
the slip length is equal to $\lambda$,
\bee
L_h= \lambda.\label{ss}
\eee

For a corrugated wall, $L_h$ has been evaluated in Ref.~\onlinecite{Cichocki_Szymczak}.\\

\section{Spherical particle in an incident flow}\label{3}
\subsection{Mutipole method}
Now we consider a new system: a spherical particle of radius $a$,  immersed in  an arbitrary incident flow ${\bf v}_0(\br)$, $p_0(\br)$ which satisfies the Stokes equations \eqref{St1}-\eqref{St}. 
Owing to the presence of this particle, 
the fluid velocity ${\bf v}_0(\br)$
is modified,
\begin{equation}
\label{55}{\bf v}={\bf v}_0+{\bf v}_1,
\end{equation}
with the disturbance ${\bf v}_1$ (called ``the reflected flow'')~\cite{Kim} 
vanishing far from the particle center, which is located at  $\br={\bf 0}$,  
\bee
{\bf v}_1(\br)
\rightarrow 0  \hspace{0.5cm} &\mbox{for}& \hspace{0.5cm} \br \rightarrow \infty.
\eee
The disturbance depends on the particle internal structure, described in terms of the  boundary conditions at the particle surface and the flow equations inside the particle. 

Calculation of the flow ${\bf v}_1$ ``reflected'' by the particle immersed in the incident flow  ${\bf v}_0$ is the basic procedure in the method of multiple reflections used to determine fluid flows in many-particle systems.~\cite{Kim}

An arbitrary incident flow ${\bf v}_0(\br)$ can be expressed as a linear combination of the elementary solutions, introduced by Lamb~\cite{Lamb}, and expressed by him in terms of the solid harmonics, which form the basic set of the solutions to the Laplace equation and are labeled by the indices
\bee
l&=&1,2,3,...,\\m&=&-l,...,l.
\eee
Each Lamb's elementary solution belongs to one of three families~\cite{Lamb}.

Linear combinations of the Lamb's elementary solutions, the multipole functions ${\bf v}^+_{lm\sigma}(\br)$, have been constructed by Felderhof and Schmitz~\cite{FS1,FS2} with the use of the irreducible representation of the group of rotations~\cite{Normand}. Three families of solutions have been labeled by an additional index,
\bee
 \sigma=0,1,2.
\eee

The multipole functions ${\bf v}^+_{lm\sigma}(\br)$  form a complete set of regular elementary solutions to the Stokes equations, therefore, 
\begin{equation}
\label{6}{\bf v}_0(\br)=\sum_{lm\sigma}c^+_{lm\sigma}{\bf v}^+_{lm\sigma}(\br).
\end{equation}
Due to linearity of the problem, it is sufficient to focus on a single term of the above series. 
Each elementary flow ${\bf v}^+_{lm\sigma}(\br)$  is ``scattered'' \cite{FJ} by the sphere centered at $\br ={\bf 0}$, and results in the following flow disturbance, 
\begin{equation}
\label{7}
{\bf v}_0(\br)\!=\!{\bf v}^+_{lm\sigma}(\br) \;\;\; \Rightarrow \;\;\;{\bf v}_1(\br)\!=\!\!\!\sum_{\sigma'=0}^2 \!
X^{(l)}_{\sigma,\sigma'} \; {\bf v}^-_{lm\sigma'}(\br),
\end{equation}
where ${\bf v}^-_{lm\sigma}(\br)$ form a complete set of such elementary solutions to the Stokes  equations, which are singular at the sphere center~\cite{FS1,FS2,FJ,Cichocki_Felderhof_Schmitz:88}. 
The advantage of choosing ${\bf v}^{\pm}_{lm\sigma}(\br)$ rather than the Lamb's solutions is that the incident flow with a given $l,m$, scattered by a spherical particle, leads to the disturbance which is characterized by the same values of $l,m$. 

The coefficients $X^{(l)}_{\sigma,\sigma'}$ are expressed in terms of the so-called scattering coefficients $A_{l\sigma}$ and $B_{l2}$, which depend on the boundary conditions at the particle surface and its internal structure, as shown in Table~\ref{tab1}.
\begin{table}[b]
 \caption{The coefficients $[-X^{(l)}_{\sigma,\sigma'}/(2l+1)]$.}\label{tab1}
\begin{tabular}{lccc}
$\;\; \sigma'$ & 0 & 1 & 2\\ 
$\sigma$ &&&\\
\hline
\hline
\\
0 & $\dfrac{2l(2l\!-\!1)}{l\!+\!1}A_{l0}$&0&$(2l\!-\!1)(2l\!+\!1)A_{l2}$
\\
1& 0&$\;l(l\!+\!1)A_{l1}\;$&0
\\
2& $(2l\!-\!1)(2l\!+\!1)A_{l2}$&0&$\dfrac{(l\!+\!1)(2l\!+\!1)^2(2l\!+\!3)}{2l}B_{l2}$
\\\\
\end{tabular}
\end{table}

The coefficients $X^{(l)}_{\sigma,\sigma'}$ (and therefore also the scattering coefficients $A_{l\sigma}$, $B_{l2}$) are evaluated from Eq. \eqref{7}, supplemented by the appropriate boundary conditions. 
The scattering coefficients $A_{l\sigma}$, $B_{l2}$ are dimensional and scale with the particle radius $a$ as
\bee
A_{l0}(a)\! \sim \!a^{2l-1},\;\; A_{l1}(a), A_{l2}(a)\! \sim \!a^{2l+1},\;\;B_{l2}(a)\!\sim \!a^{2l+3}.\nonumber\\
\eee

For example, for a hard solid sphere with the stick boundary conditions (i.e., ${\bf v}={\bf 0}$ at its surface), they are given by the following expressions,
\bee
\!\!&&{A_{l0}^{\text{hs}}}(a)={a^{2l-1}}\dfrac{2l+1}{2},\;\;\;\;\;\; {A_{l1}^{\text{hs}}}(a)={a^{2l+1}},\hspace{0.5cm} \label{A01}\\
\!\!&&{A_{l2}^{\text{hs}}}(a)={a^{2l+1}}\dfrac{2l+3}{2},\;\;\;\;\;\; {B_{l2}^{\text{hs}}}(a)={a^{2l+3}}\dfrac{2l+1}{2}.\hspace{0.9cm}\label{AB}
\eee

The scattering coefficients have been evaluated for a wide class of spherical particles; among others, for the particles with the stick-slip boundary conditions at their surfaces~\cite{FS2}, the uniformly permeable particles~\cite{JS,RFJ}, and the core-shell particles with uniformly permeable shells and solid cores~\cite{CF}.\footnote{The misprints from Ref. \onlinecite{CF} have been corrected in Ref.~\onlinecite{cssim}. }

The coefficients $A_{l\sigma}$ with $l=1,2$ have a special physical meaning, because they 
 specify how does a particle move in an arbitrary incident flow. In Ref.~\onlinecite{FJ78}, the Faxen laws have been expressed in terms of the scattering coefficients (see also Ref.~\onlinecite{CF} for the expressions using the present notation). A particle centered at $\br_k$ which undergoes the external force ${\bf F}$ and torque ${\bf T}$, and the ambient flow ${\bf v}_0$, moves with the translational and rotational velocities, ${\bf U}$ and $\oOmega$, and exerts on the fluid the stresslet ${\bf S}$, which satisfy the following relations,
\bee
{\bf F}\!\! &=&\!\! 4 \pi \eta A_{10}\left[ {\bf U} -\left(\!1\!+\!\frac{A_{12}}{10 A_{10}}\nnabla^2 \!\right){\bf v}_0(\br_k)\right],\hspace{0.4cm}\label{F1}\\
{\bf T}\!\! &=& \!\!8 \pi \eta A_{11} \left[\oOmega - \frac{1}{2} \nnabla \times {\bf v}_0(\br_k)\right],\\
S_{\alpha \beta}\!\!&=&\!\! -\frac{4\pi}{3}\eta A_{20}\left(\!1\! +\!\frac{A_{22}}{14 A_{20}}\nnabla^2\! \right)[\partial_{\alpha}v_{0\beta}(\br_k)\!+\!\partial_{\beta}v_{0\alpha}(\br_k)].\nonumber\\\label{F3}
\eee

The factors $A_{10}$, $A_{11}$ and  $A_{20}$ 
are related  to the coefficients of the particle translational, rotational and dipole mobility, respectively. In addition, with the use of the fluctuation-dissipation theorem, they also determine the translational and rotational self-diffusion, $D^t$ and $D^r$, and the intrinsic viscosity $[\eta]$,
\bee
D^t&=& \frac{k_B T}{4\pi\eta A_{10}},\label{Dt}\\
D^r&=& \frac{k_B T}{8\pi\eta A_{11}},\\
\mbox{[} \eta 
\mbox{]} \phi &=&\dfrac{A_{20}}{a^3}\phi,\label{eta}
\eee
with the volume fraction $\phi=4\pi n a^3/3$ and the particle-number concentration $n$.

\subsection{Hydrodynamic radius}
In this section, we consider a single spherical particle of radius $a$, made of the material and with the interface described in Sec.~\ref{2} by the slip length $L_h$, see e.g. Eqs~\eqref{per}, \eqref{cs} or \eqref{ss}. 
The 
slip length $L_h$ evaluated for the flat geometry~\cite{Anderson1}  determines also the slip on the surface of the spherical particle~\cite{Anderson2}, if 
\bee
L_h/a << 1.\label{sp}
\eee
The flat geometry can be understood as the limiting case of a particle with $a \rightarrow \infty$.
The normalized scattering coefficients, ${A_{l\sigma}}(a)/A_{l\sigma}^{\text{hs}}(a)$ and ${B_{l2}}(a)/B_{l2}^{\text{hs}}(a)$, are expanded in $1/a \rightarrow 0$ up to the linear terms, keeping fixed  
all the other parameters of the boundary, the interior medium and the way it extends inside from the boundary.
For the media and interfaces described by Eqs~\eqref{per}, \eqref{cs} or \eqref{ss}, what corresponds to the particle types listed after Eq.~\eqref{AB}, the same structure of the limiting expressions is obtained.
For example, in case of $\sigma=0$, the linear terms are given as 
\bee
{A_{l0}}(a) &=&
A_{l0}^{\text{hs}}(a)\left[1-(2l-1)\dfrac{L_h}{a}+{\cal O}\left(\dfrac{L_h^2}{a^2}\right)\right].\hspace{0.7cm}
\eee

The slip length $L_h$ is now used to define the hydrodynamic radius of an effective solid smooth sphere,
$a_{\text{eff}} = a-L_h.$
With this definition, we obtain for all values of $l$ and $\sigma$ the same universal formula,
\bee
{A_{l\sigma}}(a) &=&A_{l\sigma}^{\text{hs}}(a_{\text{eff}})\left[1+ {\cal O}\left(\dfrac{L_h^2}{a^2}\right)\right].\label{main}
\eee
Analogical relations hold for $B_{l2}$. Therefore, within the linear approximation, all the scattering coefficients are determined by a single parameter: the  hydrodynamic radius, no matter what are the details of their internal structure and boundary conditions. 

Experimentally, the hydrodynamic radius can be determined by diffusion, sedimentation or viscosity measurements for a single-particle, with the use of Eqs.~\eqref{Dt}-\eqref{eta}. For example,  
\bee D^t=\frac{k_BT}{6 \pi \eta a_{\text{eff}}}.
\label{hyrad}
\eee
 Within the linear approximation, the hydrodynamic radii, given by Eqs.~\eqref{Dt}-\eqref{eta}, measured in the translational diffusion, rotational diffusion and intrinsic viscosity experiments are the same; the difference is of the higher order, ${\cal O}(L_h^2/a^2)$, with $L_h
=a-a_{\text{eff}}$.\footnote{For the stick-slip boundary conditions, the difference is even smaller, i.e. ${\cal O}(\xi^2)$, where $\xi=(a-a_{\text{eff}})/a=\lambda/(a+3\lambda)$, see Ref.~\onlinecite{Felderhof76}.} Therefore, in the linear approximation, all the scattering coefficients are determined by a single  hydrodynamic radius $a_{\text{eff}}$. 

In particular, within the hydrodynamic radius model, the Faxen laws \eqref{F1}-\eqref{F3} take the form,
\bee
{\bf F}\!\! &=&\!\! 6 \pi \eta a_{\text{eff}}\left[ {\bf U} -\left(\!1\!+\!\frac{a_{\text{eff}}^2}{6}\nnabla^2 \!\right){\bf v}_0(\br_k)\right],\hspace{0.4cm}\label{F1eff}\\
{\bf T}\!\! &=& \!\!8 \pi \eta a_{\text{eff}}^3 \left[\oOmega - \frac{1}{2} \nnabla \times {\bf v}_0(\br_k)\right],\\
S_{\alpha \beta}\!\!&=&\!\! -\frac{10\pi}{3}\eta a_{\text{eff}}^3\left(\!1\! +\!\frac{a_{\text{eff}}^2}{10}\nnabla^2\! \right)[\partial_{\alpha}v_{0\beta}(\br_k)\!+\!\partial_{\beta}v_{0\alpha}(\br_k)]\nonumber\\\label{F3eff}
\eee

The 
hydrodynamic radius model is applicable to a wide class of particles with a different internal structure and boundary conditions. Droplets and bubbles (and fluid-fluid interface) are excluded, as discussed in Sec.~\ref{2}.

\section{Many-particle hydrodynamic interactions}\label{many}

\subsection{Method of reflections (scattering expansion)}
The fluid flow in many-particle systems is the sum of the ambient flow and its disturbance caused by the presence of the particles. The disturbance can be constructed by the same procedure as in Eqs.~\eqref{6}-\eqref{7}, repeated in the process of multiple reflections (or scatterings)~\cite{Kim}, for each flow ${\bf v}^-_{lm\sigma}(\br)$ reflected by a particle $k$ with its center at $\br_k$ and treated as the incident flow ${\bf v}_0(\br)$ incoming on a particle $n$. 

The  many-particle mobility can be also written as a multiple scattering series. For example, consider the translational-translational mobility matrix $\m_{ij}$, given by the relation,
\bee
{\bf U}_i = \sum_j \m_{ij} \cdot {\bf F}_j,
\eee
which determines velocity ${\bf U}_i$ of a particle $i$ in a system of particles $j$ which undergo external forces ${\bf F}_j$.

The translational-translational mobility matrix $\m_{ij}$ can be written as the following superposition of operators \cite{sed2002},
\bee
\!\!\!\!&&\!\!\m_{ij} \!=\!\m_0(i)\delta_{ij} \!+ \!(1-\delta_{ij})\m_0(i) \, {\bf Z}_0(i)\,{\bf G}(ij)\, {\bf Z}_0(j) \,\m_0(j)\!\nonumber \\
\!\!\!\!&&\!\! -\m_0(i) \, {\bf Z}_0(i)\!\!\!\!\sum_{k\ne i, k\ne j}\!\!\!\!{\bf G}(ik)\,\hat{\bf Z}_0(k)\,{\bf G}(kj)\, {\bf Z}_0(j) \,\m_0(j) + ...\label{scat}\nonumber\\
\eee
This superposition is the multiplication and summation of the corresponding matrix elements, labeled by the indices $l,\;m\;,\sigma$. For example, the matrix elements of the single-particle friction operator ${\bf Z}_0(i)$ (which depends on the particle $i$ internal structure and boundary conditions at its surface) have the form \cite{Cichocki_Felderhof_Schmitz:88},
\bee
Z_0(lm\sigma,l'm'\sigma')&=& - \eta \delta_{ll'} \delta_{mm'} X^{(l)}_{\sigma,\sigma'},\label{Z0}
\eee
and 
${\bf G}(ik)$ represents multipole elements of the Oseen tensor ${\bf T}_0({\bf r}_{ik})$~\cite{Kim}, where ${\bf r}_{ik}$ is the relative position of the centers of particles $i$ and $k$, see e.g. Ref.~[\onlinecite{sed2002,disp}] for the explicit expressions.  

The operator $\m_0(i)$ shown in Eq.~\eqref{scat}, involves only one multipole element of the single-particle translational-translational mobility, 
\bee
\mu_0(1m0,1m0)\!\! &=&
{Z_0(1m0,1m0)}^{-1}.
\eee

The interpretation of the second term in Eq.~\eqref{scat} is the following. ${\bf G}(ij)\, {\bf Z}_0(j) \,\m_0(j)$ represents the flow generated by the presence of a particle $j\ne i$ and incoming on the particle~$i$. When $\m_0(i)\, {\bf Z}_0(i)$ is applied to this flow, it results in a change of velocity of the particle $i$ after this single scattering. 

The interpretation of the third term in Eq.~\eqref{scat} is the following. Each multiple scattering from a particle $k \ne i$ adds an additional contribution to velocity of particle $i$, now with the requirement that no additional force nor torque on the particle $k$ is induced as the result of the scattering event (this requirement leads to $\hat{\bf Z}_0(k)$ rather than ${\bf Z}_0(k)$, with the definition and explicit expressions given e.g. in Ref.~[\onlinecite{m3}]).

\subsection{Rotne-Prager approximation}\label{4}
First, consider a system of 
hard solid spherical particles of radii $a$ with the stick boundary conditions, with the distance between the centers of spheres with labels $i\ne j$ denoted as $r_{ij}$. Within the Rotne-Prager approximation \cite{Rotne-Prager}, it is described by the following, positive-definite, translational-translational mobility matrix,
\bee
{\m_{ii}^{\text{RP}}}\!\! &=& \!
\frac{1}{6\pi \eta a} {\bf I},\label{RP2}
\eee
\bee
{\m_{ij}^{\text{RP}}}\!\! &=& 
\!\!\dfrac{1}{8\pi \eta} \!\left[\!\dfrac{\left({\bf I} \!+\!\hat{\br}_{ij}\hat{\br}_{ij}\right)}{r_{ij}}\! +\!\dfrac{2a^2}{3}\, \dfrac{\left({\bf I}\! - \!3\hat{\br}_{ij}\hat{\br}_{ij}\right)}{r_{ij}^3}\right]\;\; \mbox{ for } i \ne j, \;
\nonumber\\
\label{RP}\eee

These expressions correspond to the first and 
the second terms
in the scattering expansion  \eqref{scat}, 
i.e. the leading terms in the inverse inter-particle distance, up to ${\cal O}(1/r_{ij}^3)$.

The expressions \eqref{RP2}-\eqref{RP} are therefore easily generalized for particles 
with a different internal structure and boundary conditions, and in general unequal radii. 
By inserting the appropriate scattering coefficients into Table~\ref{tab1} and Eqs.~\eqref{Z0}, and taking the first and second terms in Eq.~\eqref{scat}, we obtain the generalized Rotne-Prager expressions, 
\bee
{\m_{ii}^{\text{RP}}}\!\!  &=& \! 
\frac{1}{4\pi\eta A_{10}(a_i)} 
{\bf I},\label{aeffRP}
\eee
\bee
{\m_{ij}^{\text{RP}}}\!\!\!  &=& 
\!\!\!
\dfrac{\left({\bf I} \!+\!\hat{\br}_{ij}\hat{\br}_{ij}\right)}{8\pi \eta r_{ij}}\! +\!\dfrac{1}{5}\!\left(\!\dfrac{A_{12}(a_i)}{A_{10}(a_i)}\!+\!\dfrac{A_{12}(a_j)}{A_{10}(a_j)}\!\right)\! \dfrac{\left({\bf I}\! - \!3\hat{\br}_{ij}\hat{\br}_{ij}\right)}{8\pi \eta r_{ij}^3},\nonumber \\
&&\hspace{5.3cm}\mbox{for }i \ne j.\;\;
\label{RPgen}
\eee
The scattering coefficients, which appear in the above equations, have been explicitly evaluated for a wide range of the particle models~\cite{Cichocki_Felderhof_Schmitz:88,Cichocki_Felderhof:09}. 

The question is how to relate  the generalized Rotne-Prager mobility to experiments. The self-mobility, given by Eq.~\eqref{aeffRP}, is determined e.g. from Eq.~\eqref{Dt}, with the measured value of the translational self-diffusion coefficient. Such an experiment allows to define the hydrodynamic radius according to the Stokes law, given by Eq.~\eqref{hyrad},
with separate values of $D^t_i$ and $a_{\text{eff},i}$ for each species $i$.
Therefore, 
 \bee
a_{\text{eff},i}=2A_{10}(a_i)/3.\eee

The key point is that the Rotne-Prager mobility $\m_{ij}^{\text{RP}}$, given by Eq.~\eqref{RPgen}, is well-approximated with the use of the hydrodynamic radii only. Indeed, 
taking into account the expansion of the scattering coefficients, given by Eq.~\eqref{main}, 
and using Eqs.~\eqref{A01}-\eqref{AB}, we rewrite Eq.~\eqref{RPgen}~as 
\bee
\!\!\!\!{\m_{ij}^{RP}}\!\!\!  &=& 
\!\!\!
\dfrac{1}{8\pi \eta} \!
\left[ \!
\dfrac{\left({\bf I} \!+\!\hat{\br}_{ij}\hat{\br}_{ij}\right)}{r_{ij}}\! +\!\dfrac{a_{\text{eff},i}^2\!+\!a_{\text{eff},j}^2}{3}\dfrac{\left({\bf I}\! - \!3\hat{\br}_{ij}\hat{\br}_{ij}\right)}{r_{ij}^3}
\right]\!+...
\nonumber\\
\label{RPeff}
\eee
In this way we obtain the Rotne-Prager approximation for the translational-translational mobility of the solid particles with the  hydrodynamic radii $a_{\text{eff},i}$. 

The first term in Eq.~\eqref{RPeff} 
scales as $\sim 1/r_{ij}$ and it is exact and independent of the particle internal structure and boundary conditions on its surface. The second term scales as $\sim 1/r_{ij}^3$ and it is approximate; within the ${\cal O}([(a\!-\!a_{\text{eff}})/a]^2)$
accuracy, it depends on the the particle internal structure and boundary conditions only through the hydrodynamic radius. As it follows from Eq.~\eqref{main}, the dots in Eq.~\eqref{RPeff} mean ${\cal O}([(a\!-\!a_{\text{eff}})/a]^2)$.

\vspace{-0.15cm}
\section{Effective hydrodynamic properties of suspensions}\label{direct}
\subsection{Theoretical description}
We now move on to the description of suspension transport-properties. They are given in terms of such effective coefficients as e.g. the 
sedimentation, short-time diffusion and high-frequency viscosity coefficients. 
They are evaluated by averaging the corresponding mobility coefficients for systems of many particles. For example, the translational self-diffusion coefficient is given by the following expression,
\bee
{D}_{s} &=& k_B T \frac{< \text{Tr} \m_{ii}>}{3},
\eee
where $T $ is the temperature, $k_B$ denotes the Boltzmann constant, and $\text{Tr}$ stands for the trace of a Cartesian matrix.
${D}_{s}$ is obtained by averaging coefficients of the diagonal translational-translational many-particle mobility matrix $\m_{ii}$, with respect to the directions in space, particle labels $i$, and positions of all the particles, with the equilibrium distribution.

From the analysis of the scattering coefficients, performed in the previous sections, it follows that 
the many-body-mobility 
coefficients for particles of a different internal structure and boundaries
can be approximated by the many-body-mobility 
coefficients for solid particles at the same configuration of their centers, but with the smaller  hydrodynamic radii, determined from single-particle experiments. 

The averaging of the mobility coefficients depends on the probability distribution of the particle configurations. This distribution follows from direct interactions between particles. Examples of direct interactions are  Yukawa or Debye-H\"uckel external forces, or the no-overlap requirement.  

Therefore, the appropriate approximation of suspension effective transport properties is to evaluate the mobility coefficients for the solid particles with the effective radii, and average them with the correlation functions based on the unchanged direct interactions between the particles. 

\subsection{Discussion}
The question arises how accurate is the above approximation. 
For very dense systems, the fluid flow disturbed by the presence of a close particle surface needs to be evaluated by multiple reflections, 
described by the scattering coefficients with higher values of $l$. The larger $l$, the larger is the ${\cal O}[(L_h/a)^2]$ correction, and the smaller accuracy of the  hydrodynamic radius approximation to the scattering coefficients, given by Eq.~\eqref{main}. These coefficients are needed to evaluate and average the mobility matrix for systems of close particles. Therefore, the larger volume fraction, the smaller accuracy of the  hydrodynamic radius approximation of the transport coefficients. 

Examples of the numerical estimates of the accuracy of the  hydrodynamic radius approximation to the transport coefficients can be found in Refs.~\cite{Abade_JCP:2010,CENW2011,csvir,cssim}. The direct interactions between particles considered there are given by the no-overlap condition, with the no-overlap radius equal to the geometrical radius $a$. 

Refs.~\cite{Abade_JCP:2010,CENW2011,csvir,cssim} contain reference data for the transport coefficients of suspensions made of solid particles with hydrodynamic radii $a_{\text{ff}}$, with larger no-overlap radii $a$. These results can be used to model suspensions made of particles of different internal structure and boundaries. 

The
example of the no-overlap direct interactions between permeable or core-shell particles in suspensions, discussed in Refs~\cite{Abade_JCP:2010,CENW2011,csvir,cssim}, indicates that in both cases, the typical inaccuracy of the  hydrodynamic radius approximation is very small: comparable to or less than 5\%, 
even for very condensed systems with $\phi=0.45$~\cite{Abade_JCP:2010,cssim}. 

In particular, it is worthwhile to discuss here the systems of particles of radii $c$ covered by thin polymer shells of thickness $\delta$ (the outer radii are $a=c+\delta$), with $\epsilon=\delta/c < 1$. Ref.~\cite{csvir} is focused on the dynamics of dilute suspensions of non-overlapping core-shell particles with thin, uniformly permeable shells, and the ratio $x=c\kappa$ of the inner radius $c$ to the hydrodynamic screening length $\kappa^{-1}$ of the permeable medium. In this reference, 
the virial coefficients, which correspond to pair hydrodynamic interactions, have been evaluated  and discussed.

Surprisingly, even for quite thick shells, with the relative thickness $\epsilon=\delta/c \le 0.3$ and $x \ge 10$, 
the uncertainty of the model is smaller than $0.2, 1, 2, 3\%$ for the sedimentation, translational self-diffusion, high-frequency viscosity and rotational self-diffusion, respectively. This uncertainty is significantly smaller than $\epsilon^2$. The explanation of this high precision is simple: the relative shell thickness $\epsilon$ is not the relevant parameter. The relevant parameter is the relative {\it hydrodynamic} shell thickness, $L_h/a = (a-a_{\text{eff}})/a$, with $L_h$ given by Eq.~\eqref{cs}.
$L_h/a$ is typically much smaller than $\epsilon$. For example, for $x=10$ and $\epsilon=0.3$, the relative {\it hydrodynamic} shell thickness, $L_h/a \approx 0.08-0.09$, is 3-4 times smaller than the {\it geometrical} one. 

This example illustrates the essential difference between the strategy presented in this paper in comparison to Ref.~\cite{Anderson2}. In the last reference, 
 $\epsilon$ (called there $\lambda$) was used as the expansion parameter, and the  ${\cal O}(\epsilon^2)$ corrections had to be evaluated, because they  were not small. Here, we use a much smaller expansion parameter $L_h/a$, and as the result, the terms ${\cal O}((L_h/a)^2)$ can be neglected, and no details of the particle internal structure and surface are needed, except the value of the hydrodynamic radius.

\section{Conclusions}\label{5}
From the analysis described in this paper it follows that spherical particles of radius $a$, characterized by a wide range of different internal structures and boundaries, disturb the incident flow in approximately the same way: as effective solid particles of a smaller radius $a_{\text{eff}}$, called here the  hydrodynamic radius. The relative differences 
are of the order of $[(a\!-\!a_{\text{eff}})\!/a]^2$. 

The reasoning presented here does not refer to droplets nor bubbles. It applies, for example, to such systems are permeable, core-shell or rough spherical particles, or the particles with the Navier stick-slip boundary conditions. The results obtained for the last model of the particle internal structure are especially meaningful: the Navier slip length $\lambda$ cannot be arbitrarily chosen or fitted to model realistic particles of a complex structure; its value follows from the scattering theory as $a\!-\!a_{\text{eff}}$, the difference between the particle geometrical and hydrodynamic radii.

Dynamics of suspensions of permeable, core-shell or rough spherical particles, or model particles with the Navier stick-slip boundary conditions, can be described with a reasonable accuracy by evaluating many-particle hydrodynamic mobility coefficients within the  hydrodynamic radius approximation, and averaging them with the constraint that 
direct interactions between these particles are kept unchanged. 

It has been checked that for a wide range of particle internal structures and boundaries, and the no-overlap direct interactions, deviations from the above approximation 
are very small~\cite{per,Abade_JCP:2010,cssim,csvir} and as such, hard to be detected in experiments, even for relatively dense suspensions.

\acknowledgments
M.L.E.-J. and E.W. were supported in part by the Polish NCN Grant No. 2011/01/B/ST3/05691. M.L.E.-J. acknowledges the scientific benefits from the COST Action MP1106.

\end{document}